\documentstyle[12pt]{article}
\begin{document}
\setlength{\unitlength}{1mm}

{\hfill   December 1997 }

{\hfill    Alberta-Thy 24-97}

{\hfill    hep-th/9801054} \vspace*{2cm} \\
\begin{center}
{\Large\bf Exact solution for a quantum field with $\delta$-like interaction}
\end{center}
\begin{center}
Sergey N.~Solodukhin\footnote{e-mail: sergey@phys.ualberta.ca}
\end{center}
\begin{center}
{\it Department of Physics, University of Alberta, Edmonton, Alberta, T6G 2J1, 
Canada}  
\end{center}
\vspace*{1cm}
\begin{abstract} 
A quantum field described by the field operator
$\Delta_{\bf a}=\Delta+{\bf a}\delta_\Sigma$ involving a
$\delta$-like potential is considered. Mathematically, the treatment
of the $\delta$-potential is based on the theory of self-adjoint extension
of the unperturbed operator $\Delta$. We give the general expressions for
the resolvent and the heat kernel of the perturbed 
operator $\Delta_{\bf a}$. The main attention is payed to $d=2$ 
$\delta$-potential though $d=1$ and $d=3$ cases are considered in some detail.
We calculate exactly
the heat kernel, Green's functions and the effective action
for the operator  $\Delta_{\bf a}$  in diverse dimensions and for various spaces $\Sigma$.
The renormalization phenomenon for the coupling constant $\bf a$ of $d=2$ and 
$d=3$ $\delta$-potentials  is observed. We find the non-perturbative
behavior of the effective action with respect to the renormalized coupling 
${\bf a}_{ren}$.
\end{abstract}
\begin{center}
{\it PACS number(s): 04.70.Dy, 04.62.+v}
\end{center}
\vskip 1cm
\newpage
{\it 1. Introduction}. We consider  a quantum field the dynamics of which on
 the Euclidean $d$-dimensional manifold $M^d$
is described by the field operator
\begin{equation}
\Delta_{\bf a}=\Delta +{\bf a} \delta_\Sigma
\label{1}
\end{equation}
where $\Delta$ is an unperturbed (hereafter to be Laplace operator)
operator
on $M^d$ and $\delta_\Sigma$ is $\delta$-like potential having support
on a subspace $\Sigma$. 
Our special concern is the case when 
 $M$ is a direct product of two-dimensional plane $R^2$ and $\Sigma$.
The coupling constant $\bf a$ in (\ref{1}) can be viewed as a measure
of the interaction with some background field that is concentrated
at the space $\Sigma$. If $\Sigma$ is a point the operator
(\ref{1}) describes a quantum field with point interaction.
Otherwise, the interaction is spread over the sub-space $\Sigma$.

Operators of the form (\ref{1}) arise in different fields of
physics. Our study, however, is motivated by applications in gravitational 
physics. The operators (\ref{1}) appear as a result of the non-minimal 
coupling of quantum  matter
to the gravitational background having conical singularities.
Indeed, the scalar curvature possesses a distributional behavior
at a conical singularity
\begin{equation}
{\cal R}={\cal R}^{reg}+4\pi (1-\alpha )\delta_\Sigma
\label{2}
\end{equation}
spread  over surface $\Sigma$ and having angle deficit
$\delta=2\pi (1-\alpha )$.
Therefore a non-minimal operator $\Delta_\xi=\Delta 
+\xi {\cal R}$ takes the form (\ref{1}) 
being considered on a conical space. The coupling constant
then reads ${\bf a}=4\pi (1-\alpha )\xi$.

The conical geometry arises naturally in  three dimensions \cite{1} 
as the only
 result of the 
gravitational interaction of point particles. In four and higher 
dimensions a cosmic string produces space-time which can be 
modeled by a conical space \cite{V}. 
A scalar field with the non-minimal coupling on the cosmic string
background was considered in \cite{A1}, \cite{A2}.
On the other hand, a conical singularity appears in the Euclidean approach to the black hole thermodynamics \cite{S1}.
In this context the  developing of the theory of the non-minimal coupling to
a conical background is important for understanding such outstanding 
issues as the renormalization of the black hole entropy \cite{S2},
the correspondence of various approaches to the black hole thermodynamics
\cite{S3} and a mechanism of generating  the Bekenstein-Hawking
entropy in the induced gravity \cite{F1}.

The operators of the form (\ref{1}) give  us an example of exactly solvable models
the mathematical theory of which is well known \cite{M1}, \cite{M2}
and based on
theory of self-adjoint extension of operators \cite{RS}.
However, the quantum field theoretical aspects of (\ref{1}) are not so
well developed. 
In this paper those aspects are considered systematically.
We calculate the heat kernel, Green's functions and the effective action
for the operator (\ref{1}) in diverse dimensions and for various
spaces $\Sigma$. In particular, we observe the renormalization phenomenon for 
the coupling constant $\bf a$ and find a non-perturbative behavior 
of the effective action
with 
respect to
the renormalized coupling ${\bf a}_{ren}$.

\bigskip

{\it 2. Mathematical set up}. We start with some general consideration of the
$d$-dimensional operator (\ref{1}) with the point interaction
\begin{equation}
\Delta_{\bf a}=\Delta+{\bf a}\delta (x,y)
\label{3}
\end{equation}
concentrated at $x=y$. The resolvent of operator (\ref{3}) is defined as 
solution of the equation
\begin{equation}
(-\Delta_{\bf a}-k^2)G_{{\bf a}, k^2}(x,x')=\delta (x,x')~~.
\label{4}
\end{equation}
The following Theorem is valid.

{\bf Theorem:}
{\it For the operator (\ref{3}) we find that

(i) the resolvent takes the form
\begin{equation}
G_{{\bf a},k^2} (x,x')=G_{k^2} (x,x')+{{\bf a}\over 1-{\bf a}G_{k^2}(y,y)}
G_{k^2}(y,x')G_{k^2} (x,y)~~.
\label{5}
\end{equation}

(ii) the heat kernel
$K_{\bf a}(x,x',t)$$=e^{t \Delta_{\bf a}}$ takes the form:
\begin{eqnarray}
&&K_{\bf a}(x,x',t) ={1\over 2\pi \imath } \int_{\cal C}
e^{-k^2 t}G_{{\bf a},k^2} (x,x')dk^2 \nonumber \\
&&=K(x,x',t)+{1\over 2\pi \imath } \int_{\cal C}
{{\bf a}\over 1-{\bf a}G_{k^2}(y,y)}
G_{k^2}(y,x')G_{k^2} (x,y)dk^2~~,
\label{6}
\end{eqnarray}
where  $\cal C$  is a clockwise 
contour going around the positive real axis  on the complex plane of variable $k^2$
and $K(x,x',t)$ is the unperturbed heat kernel:
$$
K(x,x',t)= {1\over 2\pi \imath } \int_{\cal C}e^{-k^2 t}G_{k^2} (x,x')dk^2~~.
$$
}
The part $(i)$ of the Theorem
can be proven by verifying  Eq.(\ref{4}) for 
the function (\ref{5}).
The part $(ii)$ follows from definition of the heat kernel.

One can see from (\ref{5})
 that there exists a bound state at $k=k_0$ corresponding to the 
pole of the resolvent and satisfying the equation $G_{k^2_0}(y,y)=1/{\bf a}$
and lying in the upper half-plane. In this case the contour $\cal C$
in (\ref{6}) contains also a circle around the pole.

So far we have not given a concrete  sense to the  
$\delta$-potential in (\ref{3}).
It should be noted that 
its action on test functions may not be  well-defined
(see  $d=2$ example below). Then, for instance, we must give a definition
to the quantity $G_{k^2}(y,y)$ appearing in (\ref{5}), (\ref{6}).
The mathematically rigorous treatment of the $\delta$-potential
in the operator (\ref{3}) requires  considering a {\it self-adjoint extension}
of the unperturbed operator $\Delta$. Particularly, it defines the action of
the $\delta$-potential on test functions and 
consists in re-formulation of
the operator (\ref{3})  in terms of the unperturbed operator $\Delta$ acting
on the space of field functions satisfying some  (dependent on ${\bf a}$)
``boundary'' condition at $x=y$.  

As an instructive example consider the operator (\ref{3}) on two-dimensional
plane $R^2$ with $\delta$-potential concentrated in the center of the polar coordinates
$(\rho , \phi )$: $\delta (x,y)={1\over 2\pi \rho }\delta (\rho )$ where
$\delta (\rho )$ satisfies the normalization condition $\int_0^{+\infty}
\delta (\rho )d\rho=1$. 
Let the unperturbed operator $\Delta$ be the two-dimensional
Laplace operator.
The self-adjoint extension of the $2d$ Laplace operator  consists
in allowing $\ln \rho$ singularity at $\rho=0$ for test field functions.
However, $\delta (\rho )\ln \rho$ is not  then well-defined. 
Thus, we need to define
 $\delta (\rho ) f (\rho , \phi )$ for a test function 
$f(\rho , \phi )$ behaving near $\rho=0$ as
$$
f(\rho , \phi )=f_0 \ln (\rho \mu ) +f_1 + O(\rho )~~,
$$
where $\mu$ is an arbitrary dimensional parameter.
Formally integrating $\Delta_{\bf a}f=\lambda f$ over small disk 
of radius $\epsilon$ around the origin and taking the limit $\epsilon
\rightarrow 0$ we get the constraint on the coefficients $f_0$ and $f_1$
\begin{equation}
2\pi f_0+{\bf a} f_1=0~~.
\label{7}
\end{equation}
This is the boundary condition which we should impose on the
field functions at the origin. It arises in a more sophisticated way  in 
\cite{M1}.
In fact, this condition is the precise formulation of the 
two-dimensional $\delta$-potential \cite{M1}.

Now we may define the {\it value at the origin} for a field
function  by just subtracting the logarithmic term: 
$f(\rho )|_{\rho=0}\equiv (1-\rho\ln (\rho\mu ) {d\over d\rho})f|_{\rho=0}$.
Then the action of the $\delta$-function is defined as follows:
$\delta (\rho )f(\rho )\equiv \delta (\rho )(1-\rho\ln (\rho\mu )
 {d\over d\rho})f|_{\rho=0}$.

For the unperturbed operator (${\bf a}=0$) the resolvent is expressed by means 
of Hankel's function
$G_{k^2}(x,x')={\imath \over 4}H^{(1)}_0(k|x-x'|)$.
Applying the above definition of the value at the origin to this resolvent 
$
({\bf a}G_{k^2}(0,0))\equiv {\bf a}(1-\rho\ln (\rho\mu ){d\over d\rho})G_{k^2}(\rho )|_{\rho=0}
$
and using the asymptote of Hankel's function 
${\imath \over 4}H^{(1)}_0(k\rho )\simeq -{1\over 2\pi}(\ln ({-\imath
k\rho \over 2})+ C)+O(\rho )$ we find  that
$
({\bf a}G_{k^2}(0,0))=-{{\bf a}\over 2\pi} (C+\ln {k\over 2\imath \mu})$,
where $C$ is the Euler constant.
Hence, applying our general formula (\ref{5}) 
to this particular case we find \cite{M1}
\begin{equation}
G_{{\bf a}, k^2}(x,x')={\imath \over 4}H^{(1)}_0(k|x-x'|)-
{\pi\over 8}{ {\bf a}\over (2\pi +{\bf a} C+{\bf a} 
\ln {k\over 2\imath \mu})}H^{(1)}_0(k|x|)H^{(1)}_0(k|x'|)
\label{8}
\end{equation}
for the resolvent of the two-dimensional operator (\ref{3}).
One can see that (\ref{8}) satisfies the condition (\ref{7}) at
$x'$ and $k$ fixed.
For the heat kernel we find according to the general expression (\ref{6})
\begin{eqnarray}
&&K_{\bf a}(x,x',t)=K(x,x',t)+K^{(1)}_{\bf a}(x,x',t)~~, \nonumber \\
&&K(x,x',t)={1\over 4\pi t}e^{-{(x-x')^2\over 4t}}~~, \nonumber \\
&&K^{(1)}_{\bf a}(x,x',t)=-{\pi \over 8} {1\over 2\pi\imath}\int_{\cal C}
dk^2 {e^{-tk^2}\over \ln {k\over \imath k_0}}H^{(1)}_0(k|x|)H^{(1)}_0(k|x'|)~~,
\label{9}
\end{eqnarray}
where $k_0=2 \mu e^{-{2\pi  \over {\bf a}}-C}$.
The resolvent (\ref{8}) has a pole at $k=\imath k_0$ corresponding to the 
bound state. Surprisingly, this bound state appears both for positive
and negative $\bf a$. However, only for positive $\bf a$ there exists a
correspondence to the unperturbed case: the bound state becomes a (non-normalizable) zero mode ($k_0\rightarrow 0$ when ${\bf a}\rightarrow +0$)
of the unperturbed operator. Contrary to this, for negative $\bf a$ the bound state tends to become infinitely heavy ($k_0\rightarrow \infty$) when
${\bf a}\rightarrow -0$. It should be noted that 
$k_0$ is that parameter
which characterizes the self-adjoint extension of the two-dimensional Laplace
operator. It is of our special interest to see how the quantum
field theoretical quantities (heat kernel, Green's function, effective action)
depend on the parameter $k_0$ ($\bf a$).

In the next Section we calculate the contour integral appearing in (\ref{9}).
Here we want to pause and make a comment regarding the usage of the
formula (\ref{9}) on the conical space $R^2_\alpha$.
In this case the angle coordinate $\phi$ changes in the limits
$0\leq \phi \leq 2\pi\alpha$, $\alpha \neq 1$. The heat kernel
on $R^2_\alpha$ is constructed via the heat kernel (\ref{9}) on
the regular plane $R^2$ by means of the Sommerfeld formula \cite{C}:
\begin{equation}
K_{R^{2}_{\alpha}}(x,x',t)=K_{R^2}(x,x',t)+{1\over 4\pi\alpha}\int_{\Gamma} \cot {w\over 2\alpha}
K_{R^2}(\phi-\phi'+w, |x|, |x'|, t)dw~~,
\label{10}
\end{equation}
where $\Gamma$ is some known contour on the complex plane.
In some sense  the Sommerfeld formula is just a way
to make $2\pi \alpha$-periodical function from a $2\pi$ periodical
function. If the $\delta$-potential in (\ref{3}) is originated from
the non-minimal coupling as explained in Introduction then we should also 
substitute for the coupling constant $\bf a$ its value 
${\bf a}=4\pi (1-\alpha )\xi$ and take into account that the angle coordinate now has period $2\pi\alpha$. The condition (\ref{7}) then becomes
$$
2\pi\alpha f_0+{\bf a}f_1=0~~.
$$

Applying the Sommerfeld formula to the heat kernel $K(x,x',t)$ (\ref{9})
of the 
unperturbed operator $\Delta$ one finds a 
non-trivial modification of the heat kernel
determined by the angle deficit at the singularity. This modification is
well studied and known explicitly \cite{C}, \cite{F2}.
Regarding the modification of the term $K^{(1)}_{\bf a} (x,x',t)$ in (\ref{9}),
which is of our interest here, we notice that $K^{(1)}_{\bf a} (x,x',t)$
is independent of the angle coordinates. This illustrates the fact that the
$\delta$-potential in (\ref{3}) describes an $s$-wave interaction.
Therefore,  the contour integral in (\ref{10}) vanishes\footnote{
$K^{(1)}_{\bf a}$ is already $2\pi\alpha$- (in fact, arbitrary-) periodical function
so the Sommerfeld formula
(\ref{10}) works trivially for it.} for $K^{(1)}_{\bf a}$.
This means that the part of the heat kernel which is due to the $\delta$-potential
occurs to be the same for the regular plane $R^2$ and the cone $R^2_\alpha$.
That is why below we consider  the operator (\ref{3}) with arbitrary
coupling $\bf a$ on a regular manifold supposing that the generalization
to a conical space is straightforward.

\bigskip

{\it 3. Calculation of the heat kernel in two dimensions.}
Here we calculate the term $K^{(1)}_{\bf a}(x,x',t)$ by performing
explicitly the contour integration in (\ref{9}).
To proceed we first change in (\ref{9}) variable of integration 
$k^2\rightarrow k$. The contour $\cal C$ then transforms to the real axis 
$(0<arg(k)<\pi )$ on the complex plane of
the variable $k$. 
Using the integral representation for a product
of  two Hankel's functions
$$
H^{(1)}_0(k|x|)H^{(1)}_0(k|x'|)=-{8\over \pi^2} \int_0^\infty d\phi
\int_0^\infty d\beta e^{\imath kZ_\phi \cosh \beta}~~,
$$
where $Z^2_\phi=|x|^2+ |x'|^2 +2|x||x'|\cosh (2\phi )$ we obtain
for $K^{(1)}_{\bf a}(x,x',t)$:
\begin{equation}
K^{(1)}_{\bf a}(x,x',t)={1\over \pi^2\imath}\int^{+\infty}_{-\infty}
{dkk\over \ln {k\over \imath k_0}} \int_0^\infty d\phi
\int_0^\infty d\beta e^{-tk^2}e^{\imath kZ_\phi \cosh \beta}~~.
\label{12}
\end{equation}
Separating in (\ref{12}) the integration over $k>0$ and $k<0$, taking into account that $k=e^{\imath \pi}|k|$ for $k<0$ and interchanging the $k$-integration
with integration over $\beta$ and $\phi$ we find  the representation for (\ref{12})
\begin{equation}
K^{(1)}_{\bf a}(x,x',t)=\int_0^\infty d\phi (A+A^*)~~,
\label{13}
\end{equation}
where
\begin{equation}
A={1\over \pi^2\imath}\int_0^\infty d\beta \int_0^\infty {dkk\over \ln {k\over 
k_0}-{\imath \pi \over 2}}e^{-tk^2}e^{\imath kZ_\phi \cosh \beta}~~.
\label{14}
\end{equation}
The function $K^{(1)}_{\bf a}(x,x',t)$ is a real quantity as 
seen from (\ref{13}). It is convenient to replace the denominator in (\ref{14}) by  an integral
$$
{1\over \ln {k\over 
k_0}-{\imath \pi \over 2}}=\imath \int_0^\infty dy e^{-{\pi\over 2}y} ({k\over k_0})^{-\imath y}~~.
$$
Then (\ref{14}) takes the form
\begin{equation}
A={k^2_0\over \pi^2}\int_0^\infty dy e^{-{\pi\over 2}y} k_0^{\imath y}
\int^\infty_0d\beta
\int^\infty_0 dkk^{1-\imath y}e^{-tk^2}e^{\imath kZ_\phi
\cosh \beta}~~.
\label{15}
\end{equation}
The integration over $k$ and $\beta$ in (\ref{15}) 
can be performed explicitly.
We skip details of the calculation just referring to helpful
formulae (3.462.1) and (7.731.1) of \cite{GR}. 
The result of the integration is
the following
$$
A={\imath\over 2\pi\sqrt{t}} Z^{-1}_{\phi} e^{-{Z^2_\phi\over 8t}}\int_0^\infty
dy {e^{-{\pi\over 2}y} \over \cosh{{\pi\over 2}y}} (\sqrt{t}k_0)^{\imath y}
 W_{{1-\imath y\over 2},  0}({Z^2_\phi\over 4t})~~,
$$
where $W_{\lambda , 0}(z)$ is Whittaker's function.
By means of the equation (\ref{13}) this gives us the desired 
expression for the heat kernel $K^{(1)}_{\bf a}$
\begin{eqnarray}
&&K^{(1)}_{\bf a} (x,x',t)={1\over \pi \imath \sqrt{t}}\int_0^\infty {d\phi \over
 Z_\phi} e^{-{Z^2_\phi\over 8t}} \int^\infty_0 {dy \over e^{\pi y}+1} \nonumber \\
&&\left((\sqrt{t}k_0)^{-\imath y}W_{{1+\imath y\over 2},  0}({Z^2_\phi\over 4t})
-(\sqrt{t}k_0)^{\imath y}W_{{1-\imath y\over 2},  0}({Z^2_\phi\over 4t})
\right)~~.
\label{16}
\end{eqnarray}
Note, that $k_0$ appears in (\ref{16}) only in the combination
 $(\sqrt{t}k_0)$. 

Some  asymptotes of the expression (\ref{16}) are easy to analyze.
Being interested
in the limit   $p=\ln \sqrt{t}k_0\rightarrow\pm \infty$
we may replace the Whittaker's function  in (\ref{16})
by its value at $y=0$: $W_{{1\over 2},0}(z)=\sqrt{z}e^{-{z\over 2}}$.
Then integrating over $\phi$ (eq.(3.337) in \cite{GR}) we find  that in this limit
the heat kernel (\ref{16}) behaves as follows
\begin{equation}
K^{(1)}_{\bf a} (x,x',t)\simeq 
-{1\over 4\pi t}e^{-{(x^2+x'^2)\over 4t}}K_0({|x||x'|\over
2t}) f(p(t))~~,
\label{17}
\end{equation}
where $K_0(z)$ is the Macdonald function, $p(t)={1\over 2}\ln ({tk_0^2})$.
The function $f(p)$ is the result of the integration over $y$
\begin{equation}
\int^\infty_0{2dy\over e^{\pi y}+1} \sin py = ({1\over p}-{1\over 
\sinh p})\equiv f(p)~~.
\label{18}
\end{equation}
It plays an important role and appears frequently in our calculation.
Remarkably, (\ref{17}) is valid in both $t\rightarrow 0$ and $t\rightarrow
+\infty$ limits.
For large $p$ the function (\ref{18}) behaves as 
$f(p)={1\over p} +O(e^{-|p|})$. Therefore, the leading term in (\ref{17})
is given by $f(t)\simeq {1\over \ln {\sqrt{t}k_0}}+ O$, where $O=O(\sqrt{t}k_0)$
for $t\rightarrow 0$ and $O=O({1\over\sqrt{t}k_0})$
for $t\rightarrow+\infty$.

Another interesting asymptotic behavior of the function (\ref{16})
occurs in the regime of large $|x|$ and $|x'|$ under $t$ fixed
so that ${|x||x'|\over t}>>1$. Then we may use 
in (\ref{16}) the large $|z|$ asymptotic
behavior of Whittaker's function $W_{\mu,0}(z)\simeq z^\mu e^{-{z\over 2}}$.
After integration over $y$ given again by (\ref{18}) we obtain that
$$
K^{(1)}_{\bf a} (x,x',t)\simeq -{1\over 2\pi t}\int_0^\infty d\phi 
e^{-{Z^2_\phi\over 8t}}~~f(\ln {2tk_0\over Z_\phi})~~.
$$
The integral over $\phi$ for large ${|x||x'|\over t}$ can be approximated by
the descent method and the result reads
\begin{equation}
K^{(1)}_{\bf a} (x,x',t)\simeq 
-{1\over 4\pi t}e^{-{(x^2+x'^2)\over 4t}}~~K_0({|x||x'|\over
2t})~~ f(\ln {2tk_0\over |x|+|x'|} )~~.
\label{19}
\end{equation}
In the leading order we make use  the 
asymptotic representation
for the Macdonald function $K_0(z)\simeq \sqrt{\pi\over 2z}e^{-z}$ and
 find that
\begin{equation}
K^{(1)}_{\bf a} (x,x',t)\simeq 
-{e^{-{(|x|+|x'|)^2\over 4t}}\over 4\sqrt{\pi t}}
~~{1\over \sqrt{|x||x'|}}~~{1\over \ln {2tk_0\over |x|+|x'|}}~~.
\label{20}
\end{equation}
This equation illustrates the 
long-range effects of  
the $\delta$-potential. They were studied in \cite{A2} from a different perspective.

Calculate now 
the trace  $Tr K^{(1)}_{\bf a}=2\pi\int_0^\infty xdxK^{(1)}_{\bf a}
(x,x,t)$ of the kernel (\ref{16}). 
For coinciding points $x=x'$ we have $Z_\phi=2x\cosh \phi$. Doing
first the $x$-integration  by means of the formula (7.621.11) of \cite{GR}
and then performing the integration over $\phi$ we find for the trace
\begin{equation}
Tr K^{(1)}_{\bf a}={1\over 2\imath}\int_0^\infty {dy\over e^{\pi y}+1}
\left({(\sqrt{t}k_0)^{-\imath y}\over \Gamma (1-\imath{y\over 2})}-
{(\sqrt{t}k_0)^{\imath y}\over \Gamma (1+\imath{y\over 2})}\right)~~,
\label{21}
\end{equation}
where $\Gamma (z)$ is Gamma function.
Note, that (\ref{21}) does not depend on sizes of the space $R^2$.
In this respect it differs from
the trace of the unperturbed
heat kernel\footnote{In presence of a conical singularity one finds
an addition $c(\alpha )$ to the heat kernel $Tr K(x,x,t)={A(R^2_\alpha )\over 4\pi t}+c(\alpha)$ 
that also does not
depend
on sizes of $R^2$ and is entirely due to the singularity \cite{F2}.
This term is of a similar nature as (\ref{21}).} 
$Tr K(x,x,t)={A(R^2)\over 4\pi t}$ which is proportional to the
(infinite) area $A(R^2)$ of $R^2$.

In the limit $p=\ln \sqrt{t}k_0\rightarrow \pm \infty$ we find
that
\begin{equation}
Tr K^{(1)}_{\bf a}\simeq -{1\over 2}f(p)=-{1\over \ln (tk^2_0)}+
{\sqrt{t}k_0\over tk_0^2-1}
~~.
\label{22}
\end{equation}
Note, that typically the trace of the heat kernel
of a differential operator expands in the Laurent series with respect to
$\sqrt{t}$. The appearance of ${1\over \ln t}$ makes  behavior
of $Tr K^{(1)}_{\bf a}$ unusual. We discuss this in some detail later.

The heat kernel (\ref{16}) and 
its trace (\ref{21}) take much simpler form  after the
Laplace transform: 
$L(x,x',s)=\int_0^\infty e^{-ts^2} K^{(1)}_{\bf a}(x,x',t)dt$ and
$L(s)=\int_0^\infty e^{-ts^2}Tr K^{(1)}_{\bf a} dt$.
Indeed, after short computation involving eqs.(3.381.4), (7.630.2)
and (6.648) of \cite{GR} and
 the eq.(\ref{18}) we  find the following simple expressions
\begin{eqnarray}
&&L(x,x',s)={1\over 2\pi} K_0(|x|s)K_0(|x'|s)f(\ln {s\over k_0})~~,
\nonumber \\
&&L(s)={1\over 2s^2}f(\ln {s\over k_0})~~,
\nonumber \\
&&
f(\ln {s\over k_0})={1\over \ln{s\over k_0}}-{2sk_0\over s^2-k_0^2}
\label{23}
\end{eqnarray}
which play a significant role in our further calculation.

Note, that we did not take into account the bound state when deriving
the expressions for the heat kernel (\ref{16}) and its trace (\ref{21}).
In particular, that is why the pole at $s=k_0$ cancels 
in the Laplace transform (\ref{23}). The contribution of the bound state
to the heat kernel is $K_{{\bf a}, bound}(x,x',t)=\pi^{-1}k_0^2 K_0(|x|k_0)
K_0(|x'|k_0) e^{tk_0^2}$ and the trace is $TrK_{{\bf a}, bound}=
e^{tk_0^2}$. However, it seems that we do not really need to add these
to the expressions (\ref{16}) and (\ref{21}). Indeed, adding 
$K_{{\bf a}, bound}$ to $K^{(1)}_{\bf a}$ we, particularly, find that
$Tr(K_{{\bf a}, bound}+K^{(1)}_{\bf a})\rightarrow 1$ if $t\rightarrow 0$.
This is not consistent with the condition  $K_{\bf a}(x,x',t)
\rightarrow \delta^2(x,x')$ if $t\rightarrow 0$ for the complete 
heat kernel.

On the other hand, the bound state can be removed (it becomes a resonance
state) if one makes an analytical continuation $k_0\rightarrow -k_0$
(we demonstrate  in Section 6
how it works for $d=1$ and $d=3$ $\delta$-potentials).
However, the integral (\ref{10}), (\ref{12}) does not change if the transform
$k_0\rightarrow -k_0$ is accompanied by changing the contour
$\cal C$ in (\ref{10}) so that $-\pi <arg (k) <0$.

All these likely mean that in two dimensions
the eigen vectors of the operator $\Delta_{\bf a}$ form the complete basis
without the bound state and we should not add it when calculating
the heat kernel.

\bigskip

{\it 4. Green's function in higher dimensions.}
Consider $(d+2)$-dimensional manifold $M$ which is a direct product of
$2$-dimensional plane $R^2$ and $d$-dimensional surface $\Sigma$.
The $\delta$-potential in operator (\ref{1}) has support
at the surface $\Sigma$. The heat kernel on 
the total space $M$ is a product of the heat kernels on $R^2$ and
$\Sigma$:
$$
K_{M}=K_{R^2}(x,x',t)~K_{\Sigma}(z,z',t)~~,
$$
where $\{x\}$ and $\{z\}$ are coordinates on $R^2$ and $\Sigma$
respectively.
The heat kernel on $\Sigma$ is convenient to represent in the form
of the Laplace transform
$$
K_{\Sigma}(z,z',t)=\int_0^\infty e^{-s^2t}g_{\Sigma}(z,z',s)ds^2~~.
$$
For particular spaces we have \\
$
i) \ \Sigma=R^1 \ ,~K_{R^1}(z,z',t)={1\over 2\sqrt{2\pi t}}e^{-{\Delta z^2\over 4t}} \ ,~
g_{R^1}(z,z',s)={1\over 2\pi s}\cos (\Delta zs)~;$ \\
$ii) \ \Sigma=R^2 \ ,~K_{R^2}(z,z',\tau , \tau',t)=
{1\over 4\pi t}e^{-{\Delta z^2+\Delta \tau^2\over 4t}} \ ,~
g_{R^2}(z,z',\tau , \tau', s)={1\over 4\pi} J_0(\sqrt{\Delta z^2+\Delta \tau^2}s)~~, $ \\
where $\Delta z=z-z'$, $\Delta \tau=\tau-\tau'$ and $J_0(z)$ 
is the Bessel function.

Green's function $G_{M}(x,x')$ on $M$ relates to the heat kernel
$K_{M}(x,x',t)$  as follows
$$
G_{M}(x,x')=\int_0^\infty dt K_{M}(x,x',t)=G_{M}^{reg}
+G_{M}^{\bf a}~~.
$$
We are interested in that part of  Green's function which is due to
the $\delta_\Sigma$-potential: $G_{M}^{\bf a}=\int_0^\infty
dt K^{(1)}_{\bf a}(x,x',t)K_\Sigma(z,z',t)$.
Equivalently, it can be written as
\begin{equation}
G_{M}^{\bf a}(x,x',z,z')=\int^\infty_0
g_\Sigma(z,z',s)L(x,x',s)2sds~~,
\label{24}
\end{equation}
where $L(x,x',s)$ is the Laplace transform (\ref{23})
of the heat kernel $K^{(1)}_{\bf a}(x,x',t)$.
The eq.(\ref{24}) gives general expression for Green's function
on a space $R^2\times \Sigma$.
For particular cases $i)~ (M=R^3)$ and $ii)~(M=R^4)$
considered above we find 
\begin{eqnarray}
&&i)~~G^{\bf a}_{R^3}(x,x',z,z')={1\over 2\pi^2}
\int_0^\infty \cos (\Delta zs) K_0(|x|s)K_0(|x'|s)f(\ln {s\over k_0})ds
~~, \nonumber \\
&&ii) ~~G^{\bf a}_{R^4}(x,x',z,z')={1\over 4\pi^2}
\int_0^\infty
J_0(\sqrt{\Delta z^2+\Delta \tau^2}s)K_0(|x|s)K_0(|x'|s)f(\ln {s\over k_0})sds
\label{25}
\end{eqnarray}
The similar expressions for Green's function in $d=3$ and $d=4$
dimensions were
found in \cite{A2}. They, however, obtain the term
${1\over \ln{s\over k_0}}$ instead of the function $f(\ln{s\over k_0})$
and, thus,  
observe a pole at $s=k_0$ suggesting interpret the integral over $s$
as a principle part integral. This pole is absent\footnote{
The $x$-dependent part of (\ref{25}) looks as an analytical continuation
of the resolvent (\ref{8}) to imaginary values $k\rightarrow \imath s$
\cite{A2}.
The term ${1\over \sinh p}$, $p=\ln{s\over k_0}$ in the function $f(p)$
then appears to serve this non-trivial continuation through
a branch point $k=0$ of the logarithmic function. I thank A.Zelnikov
for discussing this point.} (see discussion in the end of Section 3) 
in our function
$f(\ln {s\over k_0})$ and the problem of interpretation of the integral
(\ref{25}) not arises.
Recall that $k_0=\mu e^{-{2\pi\over {\bf a}}}$ so Green's functions (\ref{25}) depend on
the coupling ${\bf a}$ in a non-perturbative way.
\bigskip

{\it 5. The effective action and UV renormalization.}
Calculating the effective action $W$ for a quantum field with a
field operator (\ref{1}) on the space $M=R^2\times \Sigma$
$$
W=-{1\over 2}\int_{\epsilon^2}^\infty {dt\over t}~Tr K_M=W_{reg}+W_{\bf a}~~,
$$
where $\epsilon$ is an UV cutoff, $TrK_M=TrK_{R^2}TrK_\Sigma$,
we are again interested in the part
\begin{equation}
W_{\bf a}=-{1\over 2}\int_{\epsilon^2}^\infty {dt\over t}~TrK^{(1)}_{\bf a}
TrK_\Sigma ~~,
\label{26}
\end{equation}
which is due to the $\delta$-potential in (\ref{1}).
As in the previous Section it is convenient to use the Laplace
transform. Representing $t^{-1}TrK_\Sigma=\int^\infty_0e^{-s^2t}\tau_\Sigma(s)ds$ we find  for (\ref{26})
\begin{equation}
W_{\bf a}=-{1\over 2}\int_0^{\lambda (\epsilon )}\tau_\Sigma (s) L(s)ds~~,
\label{27}
\end{equation}
where $L(s)$ is given by (\ref{23}). It should be noted that the UV divergence
appears in (\ref{26}) when integrating 
over small $t$.  Regularized by introducing parameter $\epsilon$ in (\ref{26})
it transforms to a divergence of the integral (\ref{27}) at large values of $s$
that may be regularized by a parameter $\lambda (\epsilon )$ ($\lambda (\epsilon )\rightarrow \infty$ if $\epsilon\rightarrow  0$). For simplicity we denote that $\lambda (\epsilon )=\epsilon^{-1}$.

For particular spaces we have

$i)$ $\Sigma=R^1$: $Tr K_{R^1}={L(\Sigma)\over 2\sqrt{\pi t}}$,
$\tau_\Sigma (s)={2\over \pi}s^2 L(\Sigma)$,
where $L(\Sigma)$ is the (infinite) length of $\Sigma$.

$ii)$ $\Sigma=R^2$: $Tr K_{R^2}={A(\Sigma )\over 4\pi t}$, $\tau_\Sigma (s)={1\over 2\pi} s^3 A(\Sigma )$, where $A(\Sigma )$ is the (infinite) area
of $\Sigma$.

$iii)$ $\Sigma$  is a two-dimensional sphere $S^2$ of radius $r$: 
$Tr K_{S^2}={r^2\over t}+{1\over 3} +O({t\over r^2})$, $\tau_{S^2}(s)=
2r^2s^3+{2\over 3}s +O(1)$.

$iv)$ $\Sigma$ is a two-dimensional compact surface $\Sigma^2_g$
of radius $r$ and genus $g> 1$, its area is $A(\Sigma )=4\pi (g-1) r^2$;
$Tr K_{\Sigma^2_g}=(g-1)({r^2\over t}-{1\over 3} +O({t\over r^2}))$, 
$\tau_{\Sigma^2_g}(s)=
(g-1)(2r^2s^3-{2\over 3}s +O(1))$.

Since $\tau_\Sigma (s)$ is a polynomial with respect to $s$ the following
integrals are useful:
\begin{eqnarray}
&&S_{0}(s)=\int {ds\over s}f(\ln(s))=\ln | \ln s|-\ln {|s-1|\over s+1} ~~,
\nonumber \\
&&sS_1(s)=\int ds f(\ln s)=
Ei(\ln s)-\ln |s^2-1|~~,
\nonumber \\
&&s^2S_2(s)=\int ds s f(\ln s)=
Ei(2\ln s)-2s-\ln {|s-1|\over s+1}~~,
\label{S}
\end{eqnarray}
where $Ei(x)$ is the Exponential-Integral function defined as $Ei(x)=C+\ln |x|+\sum_{k=1}^\infty {x^k\over k k!}$. It has an useful asymptote 
$Ei(x)\simeq {e^x\over x}$ if $x\rightarrow \pm \infty$.
Applying these formulae we find  the effective action in diverse dimensions
and for various $\Sigma$. \\
$i)$ $\Sigma$ is a point, $M=R^2$:
\begin{eqnarray}
&&W_{\bf a}[R^2]=-{1\over 2}\left(S_{0}({1\over \epsilon k_0})
-S_{0}({1\over \Lambda k_0})\right)
\nonumber \\
&&\simeq
-{1\over 2} \left(\ln\ln{1\over (\epsilon k_0)}-\ln\ln (\Lambda k_0)\right)+O(\epsilon k_0)~~,
\label{28}
\end{eqnarray}
$ii)$ $\Sigma=R^1$, $M=R^3$:
\begin{eqnarray}
&&W_{\bf a}[R^3]=-{L(\Sigma)\over 2\pi\epsilon} S_1({1\over \epsilon k_0})
\nonumber \\
&&\simeq
{L(\Sigma ) \over 2\pi \epsilon \ln (k_0\epsilon )}-{L(\Sigma )
k_0\over \pi}\ln (\epsilon k_0)+O(\epsilon k_0)~~,
\label{29}
\end{eqnarray}
$iii)$ $\Sigma=R^2$, $M=R^4$:
\begin{eqnarray}
&&W_{\bf a}[R^4]=-{A(\Sigma ) \over 8\pi\epsilon^2}S_2({1\over \epsilon k_0})
\nonumber \\
&&\simeq
{A(\Sigma) \over 16\pi\epsilon^2\ln ({\epsilon k_0})}+{
A(\Sigma) k_0\over 4\pi \epsilon}+O(\epsilon k_0)~~,
\label{30}
\end{eqnarray}
$iv)$ $\Sigma=\Sigma^2_g$, $g=0, 1, 2, ...$; $M=R^2\times\Sigma^2_g$:
\begin{eqnarray}
&&W_{\bf a}[R^2\times\Sigma^2_g]=-{A(\Sigma ) \over 8\pi\epsilon^2}S_2({1\over \epsilon k_0})+B(\Sigma )\left(S_{0}({1\over \epsilon k_0})-S_{0}({1\over \Lambda k_0})\right)\nonumber \\
&&\simeq
{A(\Sigma) \over 16\pi\epsilon^2\ln ({\epsilon k_0})}+{
A(\Sigma) k_0\over 4\pi \epsilon} +B(\Sigma) \ln \ln {1\over (\epsilon k_0)}+
O(\epsilon k_0)~~,
\label{31}
\end{eqnarray}
where $B(\Sigma )={(g-1)\over 6}$ and $O(\epsilon k_0)$ are finite in the
limit $\epsilon\rightarrow 0$ terms. 
We introduced an infra-red cutoff $\Lambda$ to regularize the
integral over small $s$ and used the fact that $S_1({1\over \Lambda k_0})$ and 
$S_2({1\over \Lambda k_0})$ go to zero when $\Lambda \rightarrow \infty$
when deriving (\ref{29})-(\ref{31}).
In fact, $B(\Sigma )$ can be expressed via the scalar
curvature integrated over the surface $\Sigma$: $B(\Sigma )=-{1\over 48\pi}\int_\Sigma {\cal R}$.

We stress that the effective action  found is exact
and behaves non-perturbatively  with respect to the 
coupling $\bf a$.
On the other hand, considering the $\delta$-potential in (\ref{1})
as a perturbation one obtains \cite{S2} for the  effective action in the 
first\footnote{The next orders are ill defined \cite{A1}, \cite{S2}.} order with respect to
$\bf a$:
\begin{eqnarray}
&&W_{\bf a}^{pert}[R^2]=-{{\bf a}\over 4\pi}\ln {\Lambda\over \epsilon}~~,
\nonumber \\
&&W_{\bf a}^{pert}[R^3]=-{L(\Sigma ){\bf a}\over 4\pi^2\epsilon}~~,
\nonumber \\
&&W_{\bf a}^{pert}[M^4]=-{A(\Sigma ){\bf a}\over 32\pi^2\epsilon^2}
+B(\Sigma ) {{\bf a}\over 2\pi}\ln {\Lambda \over \epsilon}
~~.
\label{33}
\end{eqnarray}
The UV divergences in the effective action (\ref{28})-(\ref{31})
are result of the interplay of two different effects: 1) divergences
which can be absorbed in  the renormalization
of the coupling $\bf a$, and 2) standard UV divergences which are
typical for a
quantum field interacting to a fixed background. It is not hard to see
that the quantity ${1\over \ln (\epsilon k_0)}$ plays a role
of the renormalized coupling ${\bf a}_{ren}$.
Indeed, defining ${\bf a}_{ren}=-{2\pi\over \ln (\epsilon k_0)}$ we find
\begin{equation}
{\bf a}_{ren}={{\bf a}\over 1-{{\bf a}\over 2\pi}\ln ({\epsilon\mu})}
\label{34}
\end{equation}
that is standard expression for the running coupling constant in 
the quantum field theory \cite{BS}. 
The eq.(\ref{34}) can be represented as a sum
over the ``leading logarithms'':
$$
{\bf a}_{ren}\simeq  {\bf a} +{{\bf a}^2\over 2\pi}\ln (\epsilon\mu )
+O({\bf a}^3\ln^2(\epsilon\mu ))~~.
$$
Note that the value of the bound state $k_0=\epsilon^{-1}e^{-{2\pi\over {\bf a}_{ren}}}$ depends on the renormalized coupling 
${\bf a}_{ren}$ in the same way as on the bare coupling 
$\bf a$.
Using this we may re-write
the expressions (\ref{28})-(\ref{31}) for the effective action in
terms of the renormalized value ${\bf a}_{ren}$:
\begin{eqnarray}
&&W_{\bf a}[R^2]=-{1\over 2}\left(S_{0}(e^{2\pi\over {\bf a}_{ren}})-
S_{0}({\epsilon\over \Lambda}e^{2\pi\over {\bf a}_{ren}})\right)
~~,
\nonumber \\
&&W_{\bf a}[R^3]=-{L(\Sigma )\over 2\pi\epsilon } S_1(e^{2\pi\over {\bf a}_{ren}})~~,
\nonumber \\
&&W_{\bf a}[M^4]=-{A(\Sigma )\over 16\pi\epsilon^2}
S_2(e^{2\pi\over {\bf a}_{ren}})
+B(\Sigma )\left(S_{0}(e^{2\pi\over {\bf a}_{ren}})-
S_{0}({\epsilon\over \Lambda}e^{2\pi\over {\bf a}_{ren}})\right)~~.
\label{35}
\end{eqnarray}
where functions $S_0, S_1, S_2$ are defined in (\ref{S}). These functions
have  nice properties with respect to the coupling
${\bf a}_{ren}$. First of all, they well behave in the strong
coupling  limits  
${\bf a}_{ren}\rightarrow +\infty$ and ${\bf a}_{ren}\rightarrow -\infty$.
Moreover,  the both limits coincide. In fact this  is just a consequence
of the analyticity of the function $f(\ln s)$ (\ref{S}) at the point $s=1$. The limiting values are
$$
\lim_{{\bf a}_{ren}\rightarrow\pm\infty}S_0(e^{2\pi\over {\bf a}_{ren}})=\ln 2
~~, ~~\lim_{{\bf a}_{ren}\rightarrow\pm\infty}S_0({\epsilon\over \Lambda}
e^{2\pi\over {\bf a}_{ren}})=\ln\ln{\Lambda\over \epsilon}~~,
$$
$$
\lim_{{\bf a}_{ren}\rightarrow\pm\infty}S_1(e^{2\pi\over {\bf a}_{ren}})=
C-\ln 2~~,~~\lim_{{\bf a}_{ren}\rightarrow\pm\infty}S_2(e^{2\pi\over {\bf a}_{ren}})=C+
2\ln 2-2~~.
$$
Another nice property of  the functions $S_p, p=0,1,2$ appears in 
limits ${\bf a_{ren}}\rightarrow +0$ and ${\bf a_{ren}}\rightarrow -0$.
Again, it occurs that both limits coincide and are finite
$$
\lim_{{\bf a_{ren}}\rightarrow \pm 0}S_1(e^{2\pi\over {\bf a}_{ren}})\simeq 
{{\bf a}_{ren}\over 2\pi}~~,~~
\lim_{{\bf a_{ren}}\rightarrow \pm 0}S_2(e^{2\pi\over {\bf a}_{ren}})\simeq 
{{\bf a}_{ren}\over 4\pi}~~.
$$
We see from this that in the limit of small ${\bf a}_{ren}$ the divergent
$\epsilon^{-1}$ and $\epsilon^{-2}$  terms in (\ref{35})
reproduce the perturbative result (\ref{33}).
The limit ${\bf a_{ren}}\rightarrow \pm 0$
for the function $S_0(e^{2\pi\over {\bf a}_{ren}})-S_0({\epsilon\over \Lambda}e^{2\pi\over {\bf a}_{ren}})$ is  little more tricky. It is well-defined
if we assume that ${|{\bf a}_{ren}|\over 2\pi}<<(\ln {\Lambda\over \epsilon})^{-1}$. This condition means that we should take the limit of
small ${\bf a}_{ren}$ first and then consider the limit of large
ratio ${\Lambda\over \epsilon}$. Provided it is done
we find for the combination of the function $S_0$ appearing in
(\ref{35})
$$
\lim_{{\bf a_{ren}}\rightarrow \pm 0}S_0(e^{2\pi\over {\bf a}_{ren}})-S_0({\epsilon\over \Lambda}e^{2\pi\over {\bf a}_{ren}})
\simeq {{\bf a}_{ren}\over 2\pi}\ln {\Lambda\over \epsilon}
$$
that exactly reproduces the perturbative result (\ref{33}).
We, thus, see that the combination $S_0(e^{2\pi\over {\bf a}_{ren}})-S_0({\epsilon\over \Lambda}e^{2\pi\over {\bf a}_{ren}})$ is just a modification 
of the logarithmic divergence $\ln {\Lambda\over \epsilon}$ for
finite values of ${\bf a}_{ren}$.

Alternatively, we may consider the limit when ${\bf a}_{ren}$
is kept finite and ${\Lambda\over \epsilon}$ goes to infinity.
In this case the result reads
\begin{eqnarray}
&&
S_0(e^{2\pi\over {\bf a}_{ren}})-S_0({\epsilon\over \Lambda}e^{2\pi\over {\bf a}_{ren}})
\nonumber \\
&&\simeq -\ln\ln {\Lambda\over \epsilon}-\ln {|{\bf a}_{ren}|\over 2\pi}
-\ln {|e^{2\pi\over {\bf a}_{ren}}-1|\over e^{2\pi\over {\bf a}_{ren}}+1}~~,
\label{36}
\end{eqnarray}
where the first term is a non-perturbative modification of the
logarithmic UV divergence and last two terms are finite and non-perturbative
with respect to ${\bf a}_{ren}$.

The above analysis suggests that the effective action (\ref{35})
is an analytical function of the coupling constant ${\bf a}_{ren}$
which changes in the limits $-\infty \leq  {\bf a}_{ren}\leq+\infty$.
Moreover, we can identify the points ${\bf a}_{ren}=+\infty$ and
${\bf a}_{ren}=-\infty$. 
The space of the coupling constant, thus, is topologically a circle.

In order to renormalize the UV divergences which remain in the effective
action (\ref{35}) 
(in addition to the divergences in the regular part $W_{reg}$ of the action)
one as usually \cite{BD} should consider a bare action
$W_{\bf a}^B$ with bare constants $\kappa_p$ which absorb the divergences.
In four dimensions, for instance, this action may have the form
$$
W_{\bf a}^B[M^4]=\kappa_0 A(\Sigma ) S_2(e^{2\pi\over {\bf a}_{ren}})+
\kappa_1 B(\Sigma )~~,
$$
where $\kappa_0$ absorbs $\epsilon^{-2}$ divergence while the 
logarithmic $\ln\ln{\Lambda
\over \epsilon}$ divergence (\ref{36}) is absorbed in the renormalization
of $\kappa_1$. The renormalized action, thus, takes the form
\begin{equation}
 W_{\bf a}^{ren}[M^4]=\kappa^{ren}_0 A(\Sigma ) S_2(e^{2\pi\over {\bf a}_{ren}})+
B(\Sigma )(\kappa_1^{ren}+S_0(e^{2\pi\over {\bf a}_{ren}}))~~.
\label{37}
\end{equation}
The expression for $ W_{\bf a}^{ren}$ on $M^2$ and $M^3$ derives
in a similar fashion.

We complete this Section with a brief 
comment regarding the expression (\ref{34})
for the renormalized constant ${\bf a}_{ren}$. It can be re-written as follows
\begin{equation}
{\bf a}(E)={{\bf a}(M)\over 1+{{\bf a}(M)\over 2\pi}\ln{E\over M}}~~,
\label{38}
\end{equation}
where ${\bf a}(E)$ and ${\bf a}(M)$ are values of the coupling
constant $\bf a$ measured at the energy $E$ and $M$ respectively.
For a positive ${\bf a}(M)$ the behavior of the running constant (\ref{38})
reminds that of the coupling constant in QCD. It goes asymptotically
to zero (asymptotic freedom) when
energy $E$ grows. For negative ${\bf a}(M)$ we observe a different behavior:
with $E$ growing the running coupling ${\bf a}(E)$ decreases and reaches
at some critical energy $E_{cr}$
the  infinite negative value.
For $E\geq E_{cr}$ the function ${\bf a}(E)$ decreases  starting from  the $\it
positive$ infinite value till zero. The transition through the point $E=E_{cr}$, thus, is  a transition from ${\it negative}$
to ${\it positive}$ values of $\bf a$ through the infinity.
A similar behavior for $E<<E_{cr}$ happens in QED.
However, there the perturbative analysis becomes meaningless \cite{BS} 
near the point
$E=E_{cr}$ laying in the strong coupling region.
In our case we deal with an exactly solvable problem. 
So, the strong coupling regime is under control and, 
what is especially important, 
the points ${\bf a}=+\infty$ and ${\bf a}=-\infty$
can be identified. Therefore, the theory freely flows through
the transition point. Starting with
 a  ${\it negative}$ value the running coupling constant
becomes ${\it positive}$ and goes to zero for large enough energies.
The regime of $\it negative$ $\bf a$, thus, is unstable and tends to transmute
into the regime of $\it positive$ $\bf a$ which is stable and 
asymptotically free.

\bigskip

{\it 6. Other examples of $\delta$-potential.}
So far we considered the case when the $\delta$-potential in the
operator (\ref{1}) is effectively two-dimensional. This is the most 
interesting case having a number of  applications in gravitational
physics. For completeness, however, we briefly  consider here
other examples of the $\delta$-potential the solution for which can be 
done along the lines pointed in  Section 2.

\bigskip

{\it 6.1 One-dimensional $\delta$-potential.}
Let the operator (\ref{1}) be the Laplace operator on the line
$R^1$ and $\delta_\Sigma=\delta (x)$ be the Dirac delta function concentrated at the point $x=0$. As we noted in  Section 2 the 
precise formulation for the operator $\Delta_{\bf a}$ can be done in terms of
the unperturbed operator $\Delta$ acting on the functions subject
to some condition imposed at the point $x=0$. In order to find this condition we integrate the equation $-\Delta_{\bf a} f=\lambda f$ from $x=-\epsilon$
to $x=+\epsilon$. In the limit $\epsilon \rightarrow 0$ we get 
the condition
\begin{equation} 
f'(+0)-f'(-0)+{\bf a}f(0)=0~~.
\label{39}
\end{equation}
The resolvent of the unperturbed Laplace operator reads
$G_k(x,x')={\imath\over 2k}e^{\imath k|x-x'|}$. Applying our general formula
(\ref{5}) we find the expression for the resolvent of the operator $\Delta_{\bf a}$:
\begin{equation}
G_{k,{\bf a}}(x,x')={\imath\over 2k}e^{\imath k|x-x'|}
-{{\bf a}\over 2k(2k-\imath {\bf a})}e^{\imath k(|x|+|x'|)}~~,
\label{40}
\end{equation}
which satisfies the condition (\ref{39}).
The pole structure of (\ref{40}) says us that for  positive values of 
$\bf a$ there
exists a bound state at $k=k_0={\imath {\bf a}\over 2}$.

Let us consider the case of negative $\bf a$ first. Then the heat kernel
for the operator $\Delta_{\bf a}$ is given by the contour integral
(\ref{6}) which can be transformed to the integral over the real axis
of the complex plane
$$
K_{\bf a}(x,x',t)={1\over \pi \imath}\int_{-\infty}^{+\infty}
dkkG_{k,{\bf a}}(x,x') e^{-tk^2}~~.
$$
For the unperturbed operator we get the well 
known expression
$$
K(x,x',t)={1\over 2\pi }\int_{-\infty}^{+\infty}
dk e^{\imath k|x-x'|-tk^2}
={1\over 2\sqrt{\pi t}}e^{-{(x-x')^2\over 4t}}~~.
$$
For the part of the heat kernel $K_{\bf a}(x,x',t)$ that 
is due to the perturbation
after integration over $k$ we find 
\begin{equation}
K^{(1)}_{\bf a}(x,x',t)={{\bf a}\over 2\sqrt{\pi t}}
\int_0^{+\infty}dy e^{{\bf a}y} e^{-{(|x|+|x'|+2y)^2\over 4t}}~~,
\label{41}
\end{equation}
where we used that for ${\bf a}<0$
$$
{1\over 2k-\imath{\bf a}}={1\over\imath } 
\int_0^{+\infty}dye^{{\bf a}y+2ky\imath}~~.
$$
The integral representation (\ref{41}) for the kernel was derived in \cite{GS}.
The integration in (\ref{41}) performs explicitly and results
\begin{equation}
K^{(1)}_{\bf a}(x,x',t)={{\bf a}\over 4}e^{{{\bf a}^2\over 4}t}
e^{-{{\bf a}\over 2}(|x|+|x'|)}\left( 1-\Phi({|x|+|x'|\over 2\sqrt{t}}-
{{\bf a}\over 2\sqrt{t}})\right)~~,
\label{42}
\end{equation}
where $\Phi(z)={2\over \sqrt{\pi}}\int_0^z dx e^{-x^2}$ is the error function.

For the trace $Tr K_{\bf a}^{(1)}=\int_{-\infty}^{+\infty}dx
K^{(1)}_{\bf a}(x,x,t)$ we find 
\begin{equation}
TrK_{\bf a}^{(1)}={1\over 2}e^{{{\bf a}^2\over 4}t}\left( 1+\Phi({{\bf a}\sqrt{t}\over 2})\right)-{1\over 2}~~.
\label{43}
\end{equation}
It is worth noting that (\ref{43}) is function of combination 
${\bf a}\sqrt{t}$ only.
Therefore, the regime of small $\bf a$ is compatible with the regime of
small $t$.
In the limit $t\rightarrow 0$ (\ref{42}) and (\ref{43}) go to zero
that is in agreement with the condition $K_{\bf a}(x,x',t\rightarrow 0)
=\delta (x,x')$ for the total heat kernel. Similarly, in the limit $t\rightarrow \infty$ we find that $Tr K_{\bf a}^{(1)}\rightarrow 0$ for ${\bf a}
<0$. 

The expressions (\ref{42}) and (\ref{43}) have been obtained  for negative
values of the coupling
$\bf a$.  Now they can be analytically extended to positive $\bf a$.
This does not change the behavior of (\ref{42}) and (\ref{43})
for small $t$. A difference appears in the regime of large $t$. Indeed, we have $K^{(1)}_{{\bf a}>0}(x,x',t)\simeq {|{\bf a}|\over 2}e^{{{\bf a}^2\over 4}t}
e^{-{|{\bf a}|\over 2}(|x|+|x'|)}$ and $TrK_{{\bf a}>0}^{(1)}\simeq
e^{{{\bf a}^2\over 4}t}$. This is exactly the contribution of the bound state $k_0^2=-{{\bf a}^2\over 4}$. 
For the Laplace transform of (\ref{42}) we find for ${\bf a}<0$
$$
L(x,x',s)={{\bf a}\over 2s(2s-{\bf a})} e^{-s(|x|+|x'|)}
$$
that is analytical continuation of the resolvent $G_{k,{\bf a}}$ (\ref{40})
to $k=\imath s$. The trace then reads
$$
L(s)={{\bf a}\over 2s^2(2s-{\bf a})}~~.
$$
Being extended for ${\bf a}>0$, $L(x,x',s)$ and $L(s)$  have
a pole at $s={{\bf a}\over 2}$ for ${\bf a}>0$. It simply means that the 
Laplace transform does not exist for $s<{{\bf a}\over 2}$.

Calculating the effective action on space $M=R^1\times \Sigma$ we find  
the part 
$$
W_{\bf a}[M]=-{1\over 2}\int^\infty_{\epsilon^2}dtt^{-1}
TrK^{(1)}_{\bf a}TrK_\Sigma~~,~{\bf a}<0
$$  
which is due to the perturbation.

$i)$ $\Sigma=R^1$:
\begin{eqnarray}
&&W_{\bf a}[R^1\times R^1]=-{L(R^1)|{\bf a}|\over 8\sqrt{\pi}}\int^\infty_{|{\bf a}|
\epsilon /2}{d\tau\over \tau^2}\left(e^{\tau^2}(1-\Phi(\tau ))-1\right)
\nonumber \\
&&\simeq {{\bf a}\over 4\pi}L(R^1) \ln ({|{\bf a}|\epsilon
\over 2})~~.
\label{44}
\end{eqnarray}

$ii)$ $\Sigma=R^2$:
\begin{eqnarray}
&&W_{\bf a}[R^1\times R^2]=-{A(R^2)\over 32\pi}{\bf a}^2\int^\infty_{|{\bf a}|
\epsilon /2}{d\tau\over \tau^3}\left(e^{\tau^2}(1-\Phi(\tau ))-1\right)
\nonumber \\
&&\simeq -{A(R^2)\over 8\pi^{3/2}} 
{{\bf a}\over \epsilon}+{A(R^2)\over 32\pi}{\bf a}^2
\ln ({|{\bf a}|\epsilon
\over 2})~~.
\label{45}
\end{eqnarray}

$iii)$ $\Sigma=\Sigma^2_g,~g=0,1,2,..$
\begin{equation}
W_{\bf a}[R^1\times \Sigma^2_g ]\simeq -{A(\Sigma)\over 8\pi^{3/2}} 
{{\bf a}\over \epsilon}+{A(\Sigma)\over 32\pi}{\bf a}^2
\ln ({|{\bf a}|\epsilon
\over 2})~~.
\label{46}
\end{equation}
where $\epsilon$ is an UV cut-off.

For positive $\bf a$ the presence of the bound state $k_0^2=-{{\bf a}^2\over 4}$ in the spectrum of the operator $\Delta_{\bf a}$ makes the quantum theory unstable. It can be easily stabilized by adding the mass term $(\Delta_{\bf a}
-m^2)$ so that $m^2> {{\bf a}^2\over 4}$. Then the expression for the effective action reads 
$$
W_{\bf a}=-{1\over 2}\int^\infty_{\epsilon^2}dtt^{-1}
TrK^{(1)}_{\bf a}TrK_\Sigma e^{-m^2 t}~~.
$$
The mass term does not alter, however, 
the behavior of the
effective action (\ref{44})-(\ref{46})  in the UV regime though it changes
the UV finite terms. Therefore, the UV part of (\ref{44})-(\ref{46}) extends
to ${\bf a}>0$.

\bigskip

{\it 6.2 Three-dimensional $\delta$-function.}
The operator (\ref{1}) takes the form $\Delta_{\bf a}=\Delta +{\bf a}
\delta^3(x)$ and acts on functions on three-dimensional plane $R^3$. 
In order to find the boundary condition we should
impose at $x=0$ we integrate the expression $-\Delta_{\bf a}f=\lambda f$
over a ball of radius $\epsilon$ surrounding the point $x=0$. In
the limit $\epsilon\rightarrow 0$ we obtain
\begin{equation}
4\pi r^2\partial_r f|_{r=0}+{\bf a}f|_{r=0}=0~~,
\label{61}
\end{equation}
where $r$ is the radial coordinate of the spherical coordinate system
with center at $x=0$. 
The self-adjoint extension in this case consists in admitting the ${1\over r}$ singularity. So the test function
$f$ near the origin expands as $f=f_0+f_1 r^{-1}+O(r)$. The value at
the origin is defined as $f|_{r=0}\equiv f_0=\partial_r(rf)|_{r=0}$.
Then we find from (\ref{61}) the condition
$$
4\pi f_1={\bf a}f_0
$$
to be imposed at $r=0$.

The resolvent of the unperturbed three-dimensional Laplace operator
is 
$$
G_{k}(x,x')={e^{\imath k |x-x'|}\over 4\pi |x-x'|}~~.
$$  
In order to
find the resolvent of the perturbed operator according to eq.(\ref{5})
we need to calculate the quantity $G_k(0,0)$. Using the above 
definition of the value
of function at the origin we obtain that $G_k(0,0)\def \partial_r(
rG_k(0,r))|_{r=0}={\imath k\over 4\pi}$. Therefore, the perturbed resolvent 
takes the form
\begin{equation}
G_{{\bf a},k}(x,x')={e^{\imath k |x-x'|}\over 4\pi |x-x'|}
+{1\over {1\over {\bf a}}-{\imath k\over 4\pi}}~
{e^{\imath k |x|}\over 4\pi |x|}~
{e^{\imath k |x'|}\over 4\pi |x'|}~~.
\label{63}
\end{equation}
For negative $\bf a$ the spectrum contains, 
as seen from (\ref{63}), a bound state $k=k_0=-\imath 4\pi {\bf a}^{-1}$.

Calculating the heat kernel of the unperturbed Laplace operator
we find the known expression
$$
K(x,x',t)={1\over (4\pi t)^{3/2}}e^{-{(x-x')^2\over 4t}}~~.
$$
The contribution of the perturbation to the heat kernel can be isolated
from eq.(\ref{6}). It reads (for ${\bf a}>0$)
\begin{equation}
K^{(1)}_{\bf a}(x,x',t)={1\over 8(\pi t)^{3/2}}{1\over |x||x'|}
e^{{4\pi\over {\bf a}}(|x|+|x'|)}\int^\infty_{|x|+|x'|}
dzze^{-{4\pi\over {\bf a}}z}e^{-{z^2\over 4t}}~~.
\label{64}
\end{equation}
The $z$ integration  can be performed explicitly and 
we find that
\begin{eqnarray}
&&K^{(1)}_{\bf a}(x,x',t)={1\over 4\pi \sqrt{\pi t}}{1\over |x||x'|}
e^{-{(|x|+|x'|)^2\over 4t}} \nonumber \\
&&-{{\bf a}\over |x||x'|}e^{{16\pi^2\over {\bf a}^2}t}e^{{4\pi\over {\bf a}}(|x|+|x'|)}\left(1-\Phi({ |x|+|x'|\over 2\sqrt{t}}+
{4\pi\over {\bf a}}\sqrt{t})\right)~~.
\label{65}
\end{eqnarray}
One can see that the coupling $\bf a$ appears in this formula
essentially in the combination ${\bf a}^{-1}\sqrt{t}$. Therefore, the limit of
small $t$ is compatible with the limit of large $\bf a$.
Analyzing the asymptotic behavior
of (\ref{65}) we find that (\ref{65}) vanishes in the limits
of small and large $t$. Moreover, it goes to zero  in the limit
${\bf a}\rightarrow +0$ as well.

For the trace  $Tr K^{(1)}_{\bf a}=4\pi \int^\infty_0dxx^2 K^{(1)}_{\bf a}
(x,x,t)$ we find  a simple expression
\begin{equation}
Tr K^{(1)}_{\bf a}={1\over 2}e^{{16\pi^2\over {\bf a}^2}t}
\left(1-\Phi ({4\pi\over {\bf a}}\sqrt{t})\right)~~,
\label{66}
\end{equation}
which is a function of ${\bf a}^{-1}\sqrt{t}$ only.
The formulae (\ref{65}) and (\ref{66}) extend to
negative values of the coupling $\bf a$ by making use the identity
$\Phi(-z)=-\Phi(z)$. In the limit of large
$t$ then we have
$$
K^{(1)}_{\bf a}(x,x',t)\simeq {2\over |{\bf a}||x||x'|}e^{{16\pi^2\over {\bf a}^2}t}e^{{4\pi\over {\bf a}}(|x|+|x'|)}~~,~~
TrK^{(1)}_{\bf a}\simeq e^{{16\pi^2\over {\bf a}^2}t}
$$
that is due to the bound state at $k=k_0=\imath 4\pi {\bf a}^{-1}$.

For the effective action on space $M=R^3\times \Sigma$ one finds

$i)$ $\Sigma$ is a point:

\begin{equation}
W_{\bf a}=-{1\over 2}\int^{+\infty}_{4\pi\epsilon\over {\bf a}}
{d\tau\over \tau}e^{\tau^2}(1-\Phi (\tau))\simeq {1\over 2}\ln ({4\pi\epsilon\over {\bf a}})~~.
\label{67}
\end{equation}

$ii)$ $\Sigma=R^1$:

\begin{equation}
W_{\bf a}=-{L(R^1)\over 4\sqrt{\pi}}{4\pi\over {\bf a}}
\int^{+\infty}_{4\pi\epsilon\over {\bf a}}
{d\tau\over \tau^2}e^{\tau^2}(1-\Phi (\tau))\simeq
-{L(R^1)\over 4\sqrt{\pi}}\left({1\over \epsilon}+{16\sqrt{\pi}\over {\bf a}}
\ln ({4\pi\epsilon\over {\bf a}})\right)~~.
\label{68}
\end{equation}

For negative $\bf a$ we again should introduce the mass term
$(m^2>16\pi^2{\bf a}^{-2})$ in order
to stabilize the bound state. It does not affect the UV part
of the effective action.

The leading divergence in (\ref{68}) can be eliminated by adding to the 
action a term $W_B=-L(R^1)\sqrt{\pi}{\bf a}^{-1}$ with bare coupling 
${\bf a}$ and defining the renormalized coupling as follows
\begin{equation}
{\bf a}_{ren}={{\bf a}\over 1+{{\bf a}\over 4\pi\epsilon}}~~.
\label{69}
\end{equation}
It is worth noting that the renormalization (\ref{69}) as well as 
(\ref{34}) resemble the renormalization arising in $d=2,3$ non-relativistic
models with $\delta$-potential \cite{NR} (see also \cite{BF}).

\bigskip

{\it 7. General remarks.}
The treatment of the $\delta$-potential in a field operator, thus, 
includes  the following
steps: i) allow some sort of singular behavior at the ``origin''
for the field functions; ii) define value of function at
the origin and re-formulate the $\delta$-potential as a
``boundary condition'' at the origin; iii) apply the formula
(\ref{5}) to get the resolvent of the perturbed operator.
We demonstrated the efficiency of these steps when the $\delta$-potential
is effectively one-, two- and three-dimensional. Higher-dimensional
$\delta$-potentials are known to be well-defined and do not require
implementation of the procedure of self-adjoint extension.
It is interesting to note that $d=1$ and $d=3$ $\delta$-potentials are
in some way dual each other. Indeed, 
one can see that the traces (\ref{43})
and (\ref{66}) of the corresponding
heat kernels
merge (up to an additive constant) under
 the ``duality'' transformation ${\bf a}_{d=3}=-8\pi {\bf a}^{-1}_{d=2}$.
This sort of the $\delta$-potentials may arise due to  topological
defects like domain walls and vertexes. In fact, our calculation can be
generalized to include arbitrary many centers producing the $\delta$-potentials
\cite{M1} (a generalization for fermionic operators is also of 
interest \cite{f}). On the other hand, the problem we consider in this paper
may serve as a model to analyze non-perturbatively the quantum fields near
space-time singularities. As we see the quantum field theoretical quantities 
calculated above do
not possess any kind of singular behavior. 
It would be interesting to check this for
space-time singularities  which are stronger than $\delta$-function.  

\bigskip

${\bf Acknowledgments:}$ I would like to thank Andrei Zelnikov
for stimulating discussions. This work was supported by the 
Natural Sciences and Engineering Research
Council of Canada.


\begin{thebibliography} \\
\bibitem{1} G. t'Hooft, Comm. Math. Phys., {\bf 117} (1988), 685;
S. Deser, R. Jackiw,  Comm. Math. Phys., {\bf 118} (1988), 495.
\bibitem{V} A. Vilenkin, Phys. Rept. {\bf 121} (1985), 263. 
\bibitem{A1} B. Allen, A. C. Ottewill, Phys. Rev. {\bf D42} (1990), 2669.
\bibitem{A2} B. Allen, B. S. Kay and A. C. Ottewill, Phys. Rev. {\bf D53} 
(1996), 6829.
\bibitem{S1} S. N. Solodukhin, Phys. Rev. {\bf D51} (1995), 609.
\bibitem{S2} S. N. Solodukhin, Phys. Rev. {\bf D52} (1995), 7046.
\bibitem{S3} S. N. Solodukhin, Phys. Rev. {\bf D56} (1997), 4968.
\bibitem{F1} V. P. Frolov, D. V. Fursaev and A. Zelnikov,
Nucl.Phys. {\bf B486} (1997), 339; 
V. P. Frolov, D. V. Fursaev, Phys. Rev. {\bf D56} (1997), 2212.
\bibitem{M1} S. Alberverio, F. Gesztesy, R. Hoegh-Krohn, and
H. Holden, {\it Solvable Models in Quantum Mechanics} (Springer-Verlag,
New York, 1988).
\bibitem{M2} B. Kay, U. M. Studer, Comm. Math. Phys. {\bf 139} (1991), 103.
\bibitem{RS} M. Reed, B. Simon, {\it Methods of modern mathematical physics. Vol. II Fourier analysis, self-adjointness} (New York - London, Academic, 1975). 
\bibitem{C} A. Sommerfeld, Proc. Lond. Math. Soc. {\bf 28}, 417 (1897);
H. S. Carslaw, Proc. Lond. Math. Soc. {\bf 20}, 121 (1898); {\bf 8}, 365 (1910); {\bf 18}, 291 (1919); J. S. Dowker, J. Phys. {\bf A10}, 115 (1977).
\bibitem{F2} H. P. McKean, I. M. Singer, J. Diff. Geometry, {\bf 1}, 43 (1967);
             J. Cheeger, J. Diff. Geometry, {\bf 18}, 575 (1983);
 H. Donnelly,  Math. Ann. {\bf 224}, 161 (1976); P. Chang and J. S. Dowker, Nucl. Phys.
{\bf B395} 407 (1993);
 D. V. Fursaev, Class. Quant. Grav. {\bf 11}, 1431 (1994);
G. Cognola, K. Kirsten and L. Vanzo, Phys. Rev.
                  {\bf D49}, 1029 (1994).
\bibitem{GR} I. S. Gradstein, I. M. Ryzhik, {\it Tables of Integrals, Series
and Products} (Academic, New York, 1980).
\bibitem{BS} N. N. Bogoliubov, D. V. Shirkov, {\it Introduction to The Theory
of Quantized Fields}  (New York, Wiley, 1980).
\bibitem{BD} N. D. Birrell and P. C. W. Devies, {\it Quantum fields in curved space} (Cambridge University Press, Cambridge, 1982).
\bibitem{GS} B. Gaveau, L. S. Schulman, J. Phys. {\bf A19}, 1833 (1986).
\bibitem{NR} C. Thorn, Phys. Rev. {\bf D19}, 639 (1979);
R. Jackiw, {\it in M. A. Beg Memorial Volume}, edited by A. Ali and 
P. Hoodbhoy (World Scientific, Singapore, 1991);
D. K. Park, J. Math. Phys. {\bf 36}, 5453 (1995).
\bibitem{BF} F. A. Berezin, L. D. Faddeev, Soviet. Math. Dokl. {\bf 2}, 372 (1961).
\bibitem{f} V. Korepin, Theor. Math. Phys. {\bf 41},  953 (1979);
P. de Sousa Gerbert and R. Jackiw, Comm. Math. Phys. {\bf 129}, 229 (1989).




\end{thebibliography}
\end{document}